\documentstyle[12pt]{article} \hoffset -0.5in \textwidth 6.5in \textheight
9.00in  \parskip 7pt
\begin{document}
\begin{titlepage}

\rightline{\vbox{\halign{&#\hfil\cr
&CERN-TH.6987/93\cr
&ANL-HEP-CP-93-63\cr
&August, 1993\cr}}}
\vspace{0.3in}

\begin{center}
{\Large\bf
Bottom Quark Cross Sections at Collider and\\
Fixed-Target Energies at the SSC and LHC}\footnote{
To be published in the Proceedings of the Workshop on B Physics at
Hadron Colliders, Snowmass, Colorado, June 1993.

\vspace{0.4in}

\normalsize
\leftline{\vbox{\halign{&#\hfil\cr
&CERN-TH.6987/93\cr
&August, 1993\cr}}}
}

\vskip   .3in
\large Edmond L. Berger$^{a,b}$ and Ruibin Meng$^a$  \\
\vskip .2in
$^a$High Energy Physics Division
\\Argonne National Laboratory, Argonne, IL 60439, USA\\
$^b$CERN, Geneva, Switzerland\\
\end{center}

\begin{abstract}
Calculations of inclusive cross sections for the production of bottom
quarks in proton-proton collisions are presented as a function of
energy, transverse momentum, and Feynman $x_F$ for values of $\sqrt{s}$ from
$100~$GeV to $40~$TeV.  In addition, we provide simple parametrizations
of our theoretical results that should facilitate
estimates of rates, acceptances, and efficiencies of proposed new detectors.
\end{abstract}

\end{titlepage}

Calculations of heavy flavor cross sections at the planned energies of future
hadron colliders assist in the design of experiments and in the evaluation
of the merits of various options, such as
experiments in fixed target modes and/or with detection concentrated
in forward or central regions of phase space.  In this paper, calculations
are presented of inclusive cross sections for bottom quark production at
energies from $\sqrt{s}= 100~$GeV to $40~$TeV.  For several energies, cross
sections are displayed as functions for transverse momentum ($p_T$) for
selected values of Feynman $x_F$.  In addition, we provide simple analytic
parametrizations of our theoretical predictions that should make our results
easy to use in studies of expected acceptances and efficiencies of proposed new
detectors.  The theoretical computations are based on next-to-leading
order QCD hard-scattering cross sections$^{1,2}$
and the latest two-loop evolved parton densities obtained from a
global fit of data from deep-inelastic lepton scattering and
other reactions$^3$.
The work presented here is an update of an earlier publication$^4$
to which we refer for the theoretical
formalism and summary of its limitations.

The heavy quark inclusive cross section in perturbative QCD is obtained as a
convolution of parton densities $f_{i/h}(x,\mu)$ with a hard-scattering
cross section $\hat{\sigma}_{ij}(\hat{s},M_Q,\mu,\alpha_s (\mu))$.  The heavy
quark mass is $M_Q$; $\hat{s}$ is the square of the parton-parton
center-of-mass energy,
$\hat{s} = x_1x_2s$; and $\mu$ is the renormalization/factorization scale that
serves to separate long- and short-distance effects.
$$
\sigma (s, M^2_Q) =
\int\limits^1_0 dx_1 \int\limits^1_0 dx_2
f_{i/h_1} (x_1, \mu) f_{j/h_2}
(x_2, \mu) \hat{\sigma}_{ij} (\hat{s}, M_Q, \mu, \alpha_s(\mu)).
$$
For the total cross sections reported here and for the
cross sections differential in $p_T$, we use the scale choice
$\mu = \sqrt{M_b^2+p_T^2}$.  We adopt $M_b = 4.75~$GeV.
Poor knowledge of the gluon density
is a principal source of
uncertainty for predictions of bottom quark cross sections at collider
energies.  In earlier work, we showed that measurements of the bottom
quark cross sections by the CERN UA1 and FNAL CDF collaborations provide
valuable constraints on the gluon density at intermediate values of Bjorken
$x$$^5$.
For LHC and SSC energies,
the values of $x$ of interest extend over the range $10^{-6}$ to $10^{-2}$,
or so.
To explore the uncertainties associated with relative ignorance of the gluon
density in this range, we adopt for part of our study the ``singular" and
``regular" gluon density parametrizations of the MRS collaboration.  These
are denoted, respectively, MRS $D'_-$ and MRS $D'_o$.

In Fig. 1 we show calculations of the inclusive cross
section $\sigma (pp \rightarrow b\overline{b}X)$
obtained from an integration of the inclusive
yield over all rapidity and transverse momentum.
For the energy range
$100\ GeV < \sqrt{s} < 2\ TeV$ shown in Fig.~1 (a), the singular and regular
parametrizations of the gluon density yield similar results.  The differences
become increasingly significant as $\sqrt{s}$ increases above
$\sqrt{s} = 2\ TeV$, as shown in Fig.~1 (b).

A $20\ TeV$ proton beam incident on a fixed proton target provides a
center-of-mass energy $\sqrt{s} = 200\ GeV$.  For this SSC fixed-target option,
we calculate $\sigma(b\overline{b} X) = 1.27\ \mu b$ for the singular set
$D'_-$ and $1.56\ \mu b$ for the regular set $D'_o$.  Including
uncertainties associated with variations of the choice of scale $\mu$
and b quark mass $M_b$, we estimate
$$
\sigma (b\overline{b} X, \sqrt{s} = 200\ GeV) = 1.0\ {\rm to}\ 2.0 \mu b
$$

At $\sqrt{s} = 40\ TeV$, the calculations yield
$\sigma(b\overline{b} X) \simeq 0.45\ mb$ for set $D'_o$ and nearly $2\ mb$ for
set $D'_-$.  However, at this energy, gluon resummation effects$^6$
are expected to be very
significant, providing enhancement factors$^6$ of $\sim 4$ for a regular
gluon starting distribution $(xG(x) \rightarrow$ constant) and $\sim 1.5$ for a
singular distribution $(xG(x) \sim x^{-{{1} \over {2}}})$.  Using these
factors, we may multiply our $O(\alpha^3_s)$ results at $\sqrt{s} = 40\ TeV$,
obtaining $\simeq 1.8 mb$ for set $D'_-$ and $\simeq 3 mb$ for set $D'_o$.
Including estimates of other uncertainties, we quote
$$
\sigma (b\overline{b} X, \sqrt{s} = 40\  TeV) = 1\ {\rm to}\ 3 mb
$$
Similar reasoning leads to an estimate appropriate at the LHC energy
$\sqrt{s} = 15.4\ TeV$:
$$
\sigma (b\overline{b} X, \sqrt{s} = 15.4\ TeV) = 0.5\ {\rm to}\ 0.9 mb.
$$

In Fig. 2, we present the QCD order $\alpha_s^3$ differential cross
section $d\sigma/dx_F dp_T^2$ as a function of $p_T$ for the SSC collider
energy, for three values of Feynman $x_F$.  For these results, we use the
regular gluon density MRS $D'_o$.  In our earlier paper$^4$, we showed
that the influence of the more singular gluon density is felt most strongly
at small $p_T$ ($p_T \le 25\ GeV$) at
$\sqrt{s}= 40\ TeV$.
Our theoretical
results are shown as the solid curves in Fig. 2.  Notable in Fig. 2 is the
dramatic decrease in the cross section at $p_T = 0$ by more than four orders
of magnitude when $x_F$ is increased from 0. to 0.25.  Comparison of
Figs. 2 (b) and 2 (c) shows that this large drop is
followed by a less remarkable decrease by a factor of 30 or so as $x_F$ is
increased from 0.25 to 0.5.  At small $x_F$ and small $p_T$, the cross
section is sensitive to the small $x$ behavior of the gluon density.  In our
previous paper, we provided calculations of rapidity and pseudo-rapidity
distributions$^4$.

The theoretical results shown in Fig. 2 may be fitted with a fairly simple
analytic expression.  The form we adopted is
$$
{{d\sigma} \over {dp_T^2 dx_F}} = {{1} \over {(p_T^2+m_b^2)^2}}
\exp{(A+B\,p_T)}
$$

We treat quantities $A$,$B$, and $m_b$ as three free
parameters whose values vary with $\sqrt{s}$ and $x_F$.  The histograms in
Fig. 2 show the results of our fits to the theoretical calculations.  Fitted
values of the three parameters are provided in Table I.a.  The fitted value of
our parameter $m_b$ is close to the value we used for the
physical bottom quark mass $M_b$ in our calculation, to be anticipated since
$<p_T>$ is approximately $M_b$, but no great significance should be attached
to the fact that our fitted $m_b$ varies somewhat with $x_F$ and $\sqrt{s}$.
The good agreement of the simple fit with the full theoretical calculation
should make the fitted expression useful for estimates of
rates, acceptances, and efficiencies.

\vglue 0.25in
\centerline{Table I.a Fitted Parameters for $d\sigma/dp_T^2/dx_F$
for: $0 < p_T < 50~$GeV at SSC Collider Energy.}
\begin{center}
\begin{tabular}{|l|c|c|c|}
\hline
$x_F$  &  $m_b$ (GeV)  &  $A$  &  $B(GeV^{-1})$  \\
\hline
0.0   &  5.00  &  15.4  &  -0.0559  \\
0.25  &  6.16  &  5.84  &  -0.0152  \\
0.50  &  6.60  &  2.55  &  -0.0145  \\
\hline
\end{tabular}
\end{center}

\vglue 0.25in
\centerline{Table I.b Fitted Parameters for $d\sigma/dp_T^2/dx_F$
for: $0 < p_T < 20~$GeV at $\sqrt{s}=200\ GeV$.}
\begin{center}
\begin{tabular}{|l|c|c|c|}
\hline
$x_F$  &  $m_b$ (GeV)  &  $A$  &  $B(GeV^{-1})$  \\
\hline
0.0   &  5.53  &  7.16  &  -0.363  \\
0.25  &  6.34  &  4.06  &  -0.274  \\
0.50  &  6.14  &  0.383 &  -0.246  \\
\hline
\end{tabular}
\end{center}

\vglue 0.25in
\centerline{Table I.c Fitted Parameters for $d\sigma/dp_T^2/dx_F$
for: $0 < p_T < 30~$GeV at LHC Collider Energy.}
\begin{center}
\begin{tabular}{|l|c|c|c|}
\hline
$x_F$  &  $m_b$ (GeV)  &  $A$  &  $B(GeV^{-1})$  \\
\hline
0.0   &  5.35  &  14.2  &  -0.0873  \\
0.25  &  5.96  &  5.58  &  -0.0318  \\
0.50  &  6.81  &  2.46  &  -0.0376  \\
\hline
\end{tabular}
\end{center}

\vglue 0.25in
\centerline{Table I.d Fitted Parameters for $d\sigma/dp_T^2/dx_F$
for: $0 < p_T < 15~$GeV at $\sqrt{s}=120\ GeV$.}
\begin{center}
\begin{tabular}{|l|c|c|c|}
\hline
$x_F$  &  $m_b$ (GeV)  &  $A$  &  $B(GeV^{-1})$  \\
\hline
0.0   &  6.76  &  6.80  &  -0.516  \\
0.25  &  7.95  &  4.58  &  -0.440  \\
0.50  &  7.74  &  1.02  &  -0.408  \\
\hline
\end{tabular}
\end{center}

In Fig. 3 and Table I.b, we provide analogous results for the SSC fixed target
energy of $\sqrt{s}= 200\ GeV$.  The $x_F$ dependence is less dramatic at this
energy.

In Fig. 4 and Table I.c, we present results for the LHC collider energy of
$\sqrt{s}= 15.4\ TeV$.  The $x_F$ dependence at this energy shows a steep
decrease from $x_F$ = 0 to 0.25, followed by a more gradual decrease from
0.25 to 0.5, as we saw above at $\sqrt{s}= 40\ TeV$.  Fitted values for the
LHC cross section at $\sqrt{s}= 120\ GeV$ are provided in Table I.d

We acknowledge valuable communications with Stanley Wojcicki.
We are grateful to
W.~James Stirling for providing a copy of the
Martin, Roberts, and Stirling parton densities.
This work was supported in part by the U.S. Department of
Energy, Division of High Energy Physics, Contract W-31-109-ENG-38.

\section*{References}

\begin{enumerate}
\item\label{1}
P. Nason, S. Dawson, and R. K. Ellis, Nucl. Phys. {\bf B303}, 607
(1988); {\bf B327}, 49 (1989).
\item\label{2}
W. Beenakker, H. Kuijf, W. L. van Neerven, and J. Smith, Phys. Rev.
{\bf D40}, 54 (1989); W.~Beenakker, W.~L.~van Neerven, R.~Meng, G.~Schuler, and
J.~Smith, Nucl. Phys. {\bf B351}, 507 (1991).
\item\label{3}
A. D. Martin, R. G. Roberts, and W. J. Stirling,
Phys. Lett. {\bf B306}, 145 (1993), Erratum-ibid. {\bf B309}, 492 (1993).
\item\label{4}
E. L. Berger, and R. Meng, {\em Phys. Rev.} {~\bf D46},
169 (1992).
\item\label{5}
E. L. Berger and R. Meng, {\em Phys.Lett.} {~\bf B304}, 318 (1993);
E. L. Berger and R. Meng, in the {\em
Proc. DPF92}, Fermilab, 1992,
edited by C. Albright, P. Kasper, R. Raja, and
J. Yoh, vol. 2, p. 954; E. L. Berger, R. Meng, and J.-W. Qiu,
in the {\em Proc. XXVI Int. Conf. on High Energy Physics}, Dallas, 1992,
edited by J. R. Sanford, vol. 1, p. 853;
E. L. Berger, R. Meng, and W.-K. Tung, {\em Phys. Rev.}
{~\bf D46}, 1859 (1992).
\item\label{6}
J.C. Collins and R.K. Ellis,
{\em Nucl. Phys.} {\bf B360}, 3 (1991);
S. Catani, M. Ciafaloni, and F. Hautmann,
{\em Nucl. Phys.} {\bf B366}, 135 (1991);
E.M. Levin {\it et al.}, {\em Sov. J. Nucl. Phys.} {\bf 54}, 867 (1991).
\end{enumerate}

\section*{Figure Captions}
\begin{description}

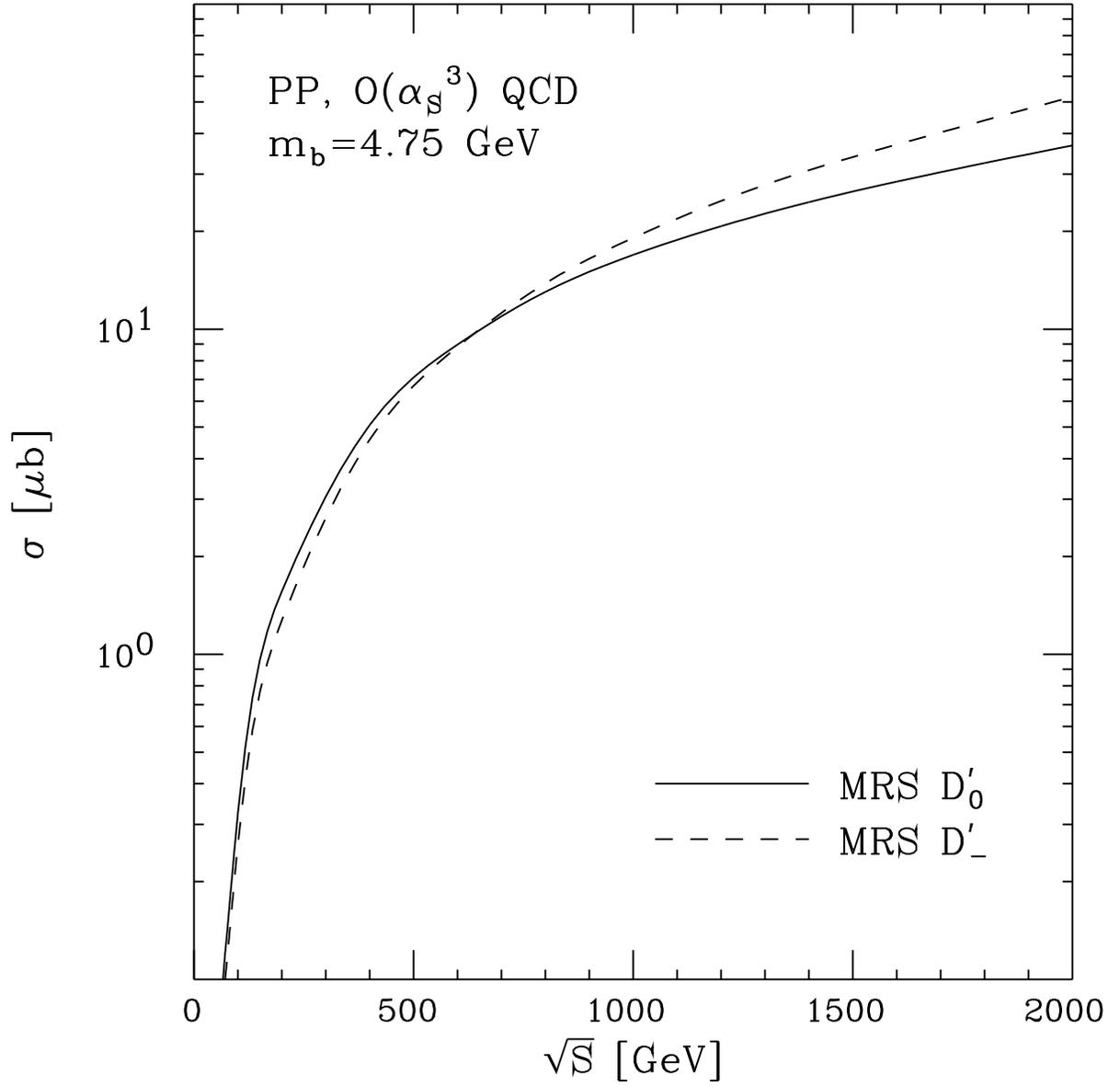
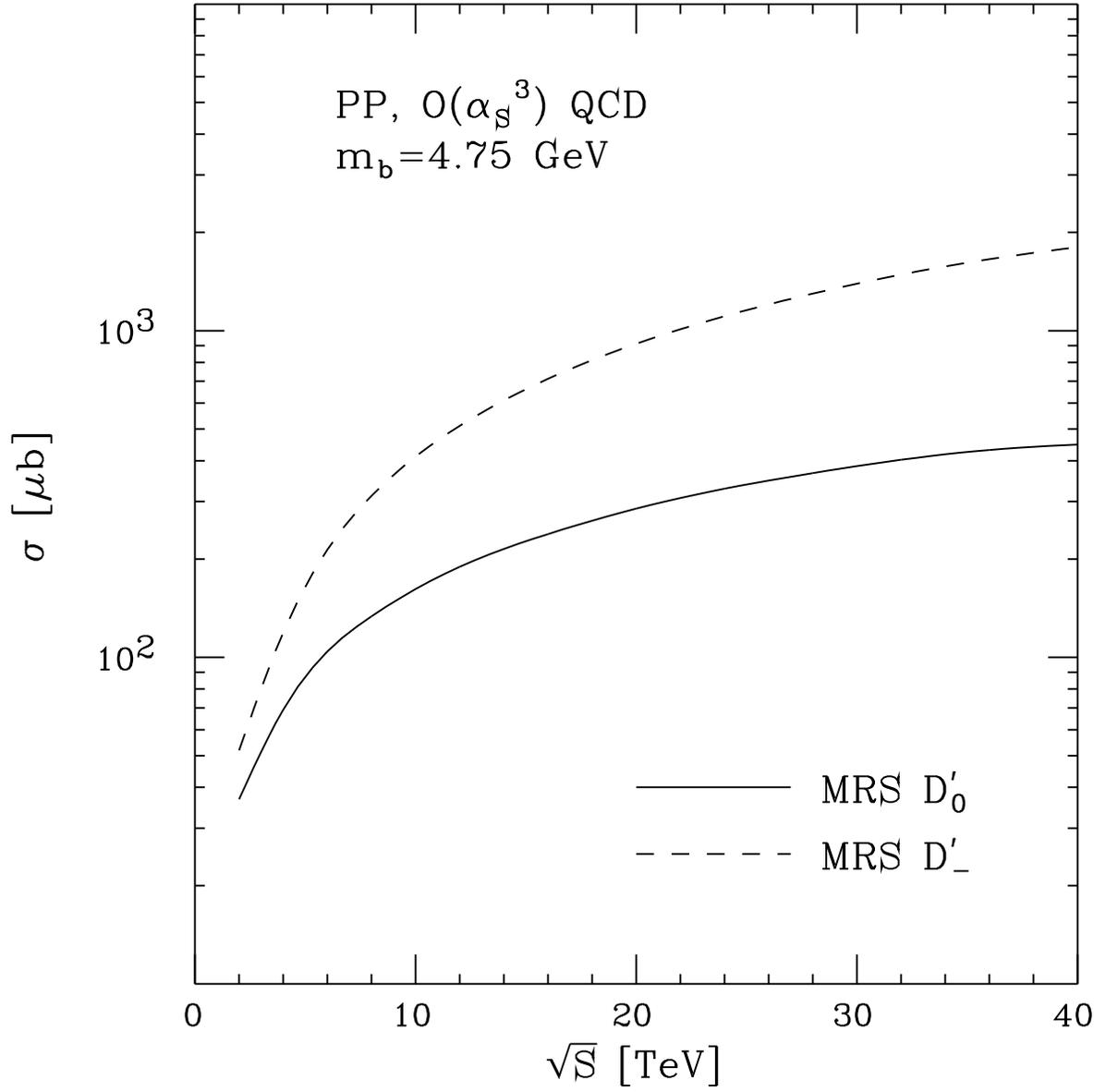
\item[Fig.  1.]
a) The calculated cross section $\sigma(pp\rightarrow b\bar bX)$ at order
$\alpha_s^3$ in QCD is shown as a function of $\sqrt{s}$ for $100 <
\sqrt{s} < 2000$ GeV.  The solid line is obtained from the parton
densities of MRS $D'_o$ (``regular'' gluon), and the dashed curve
from the MRS $D'_-$(``singular'' gluon).  b) As in (a), but
for $\sqrt{s} \ge 2~$TeV.

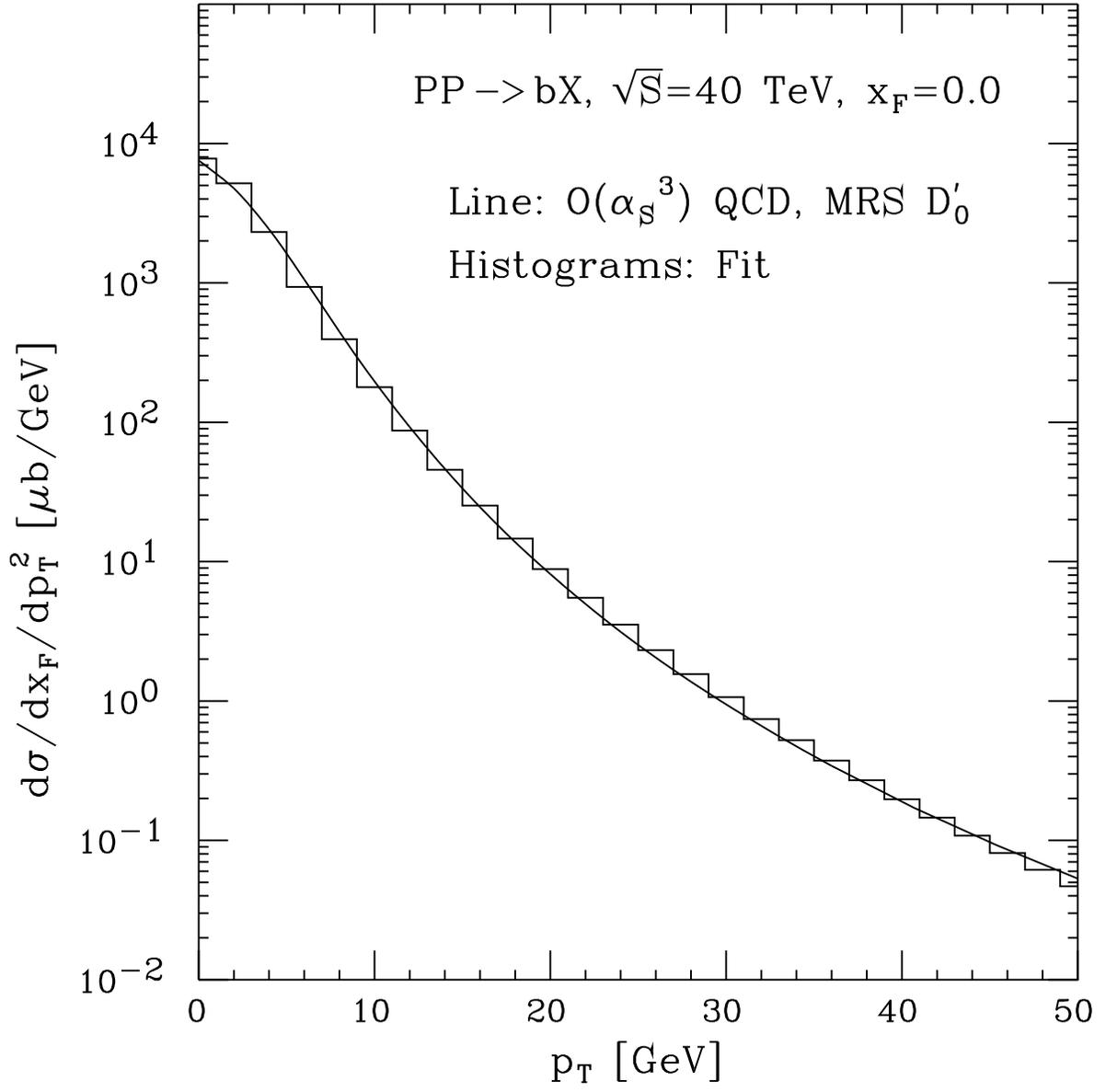
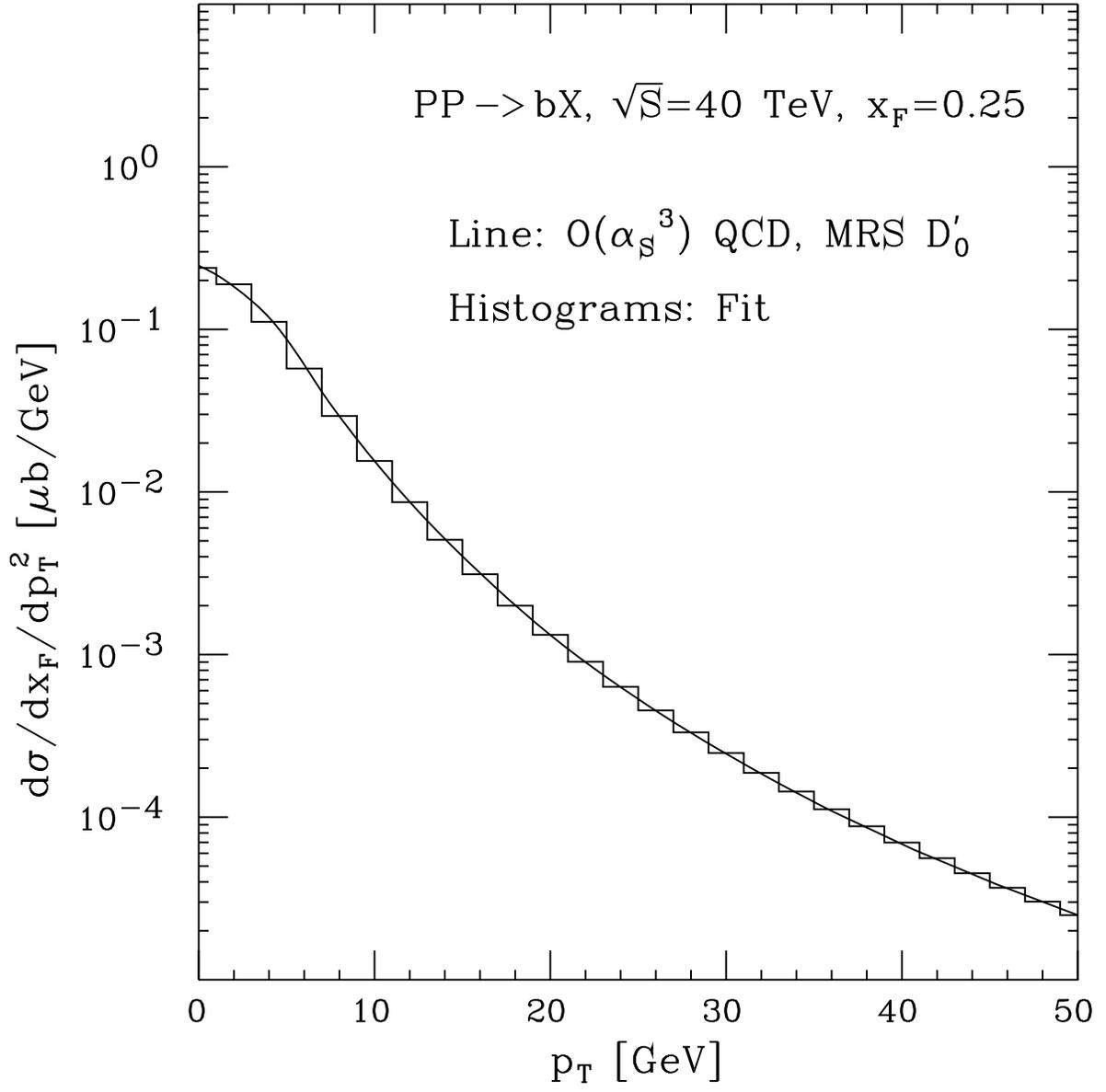
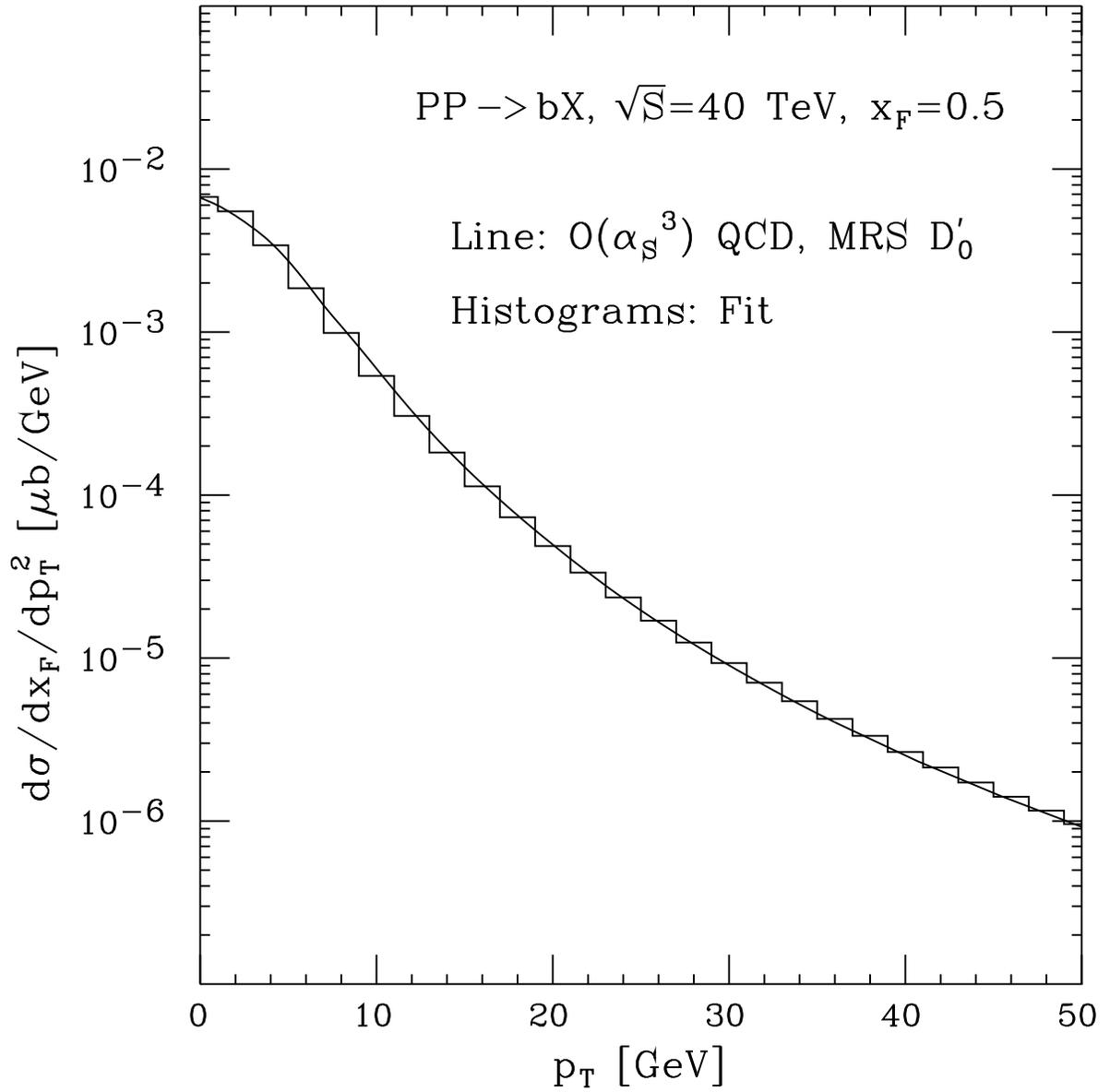
\item[Fig.  2.]
The calculated order $\alpha_s^3$ QCD differential cross section
$d\sigma/dx_F/dp_T^2$ vs. $p_T$ for the SSC collider energy $\sqrt{s}=40~$TeV
is shown (solid curves) for
(a)$x_F=0$; (b)$x_F=0.25$; (c)$x_F=0.50$.  Phenomenological parametrizations
fitted to the theoretical calculations are presented as the histograms.

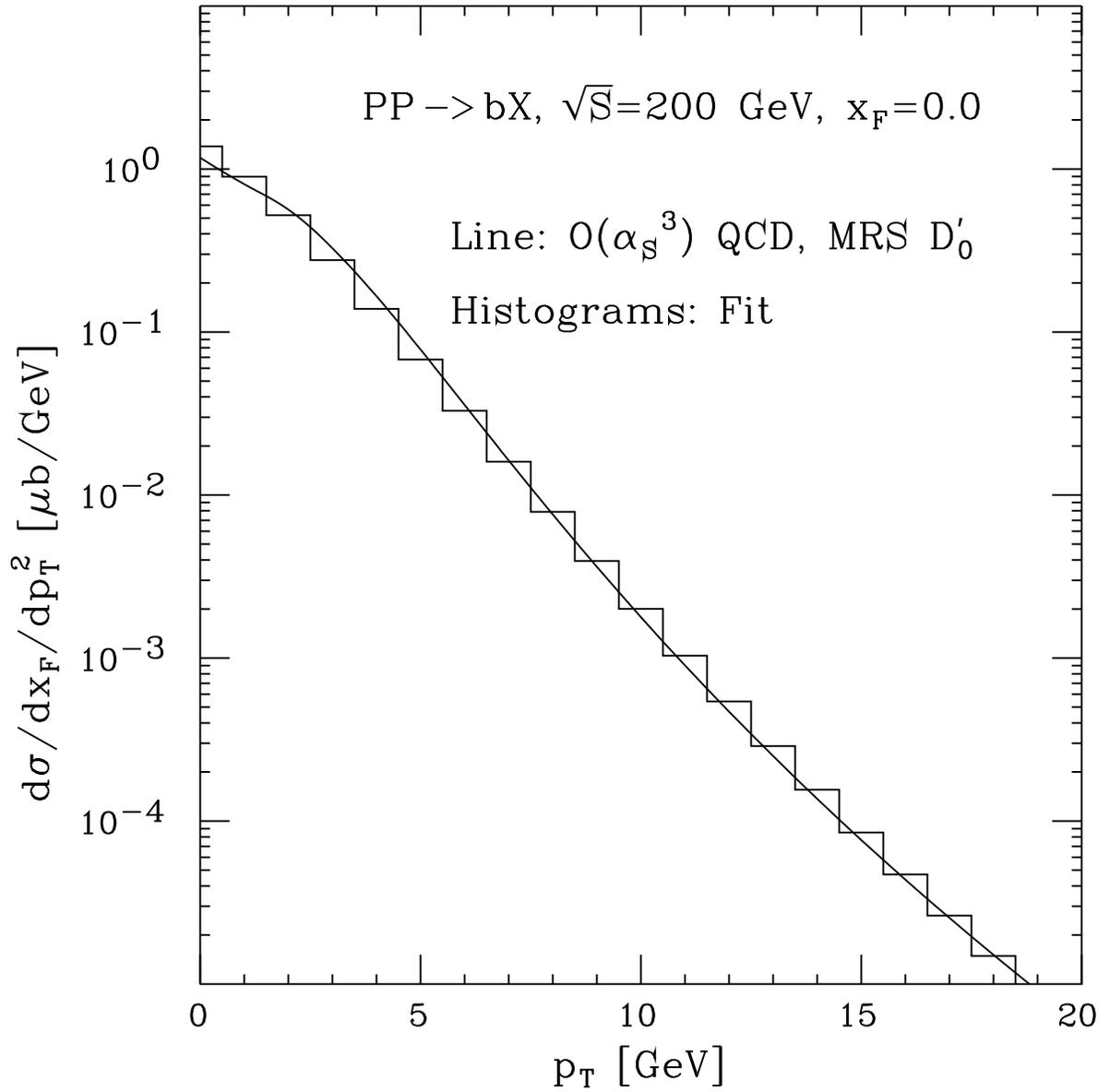
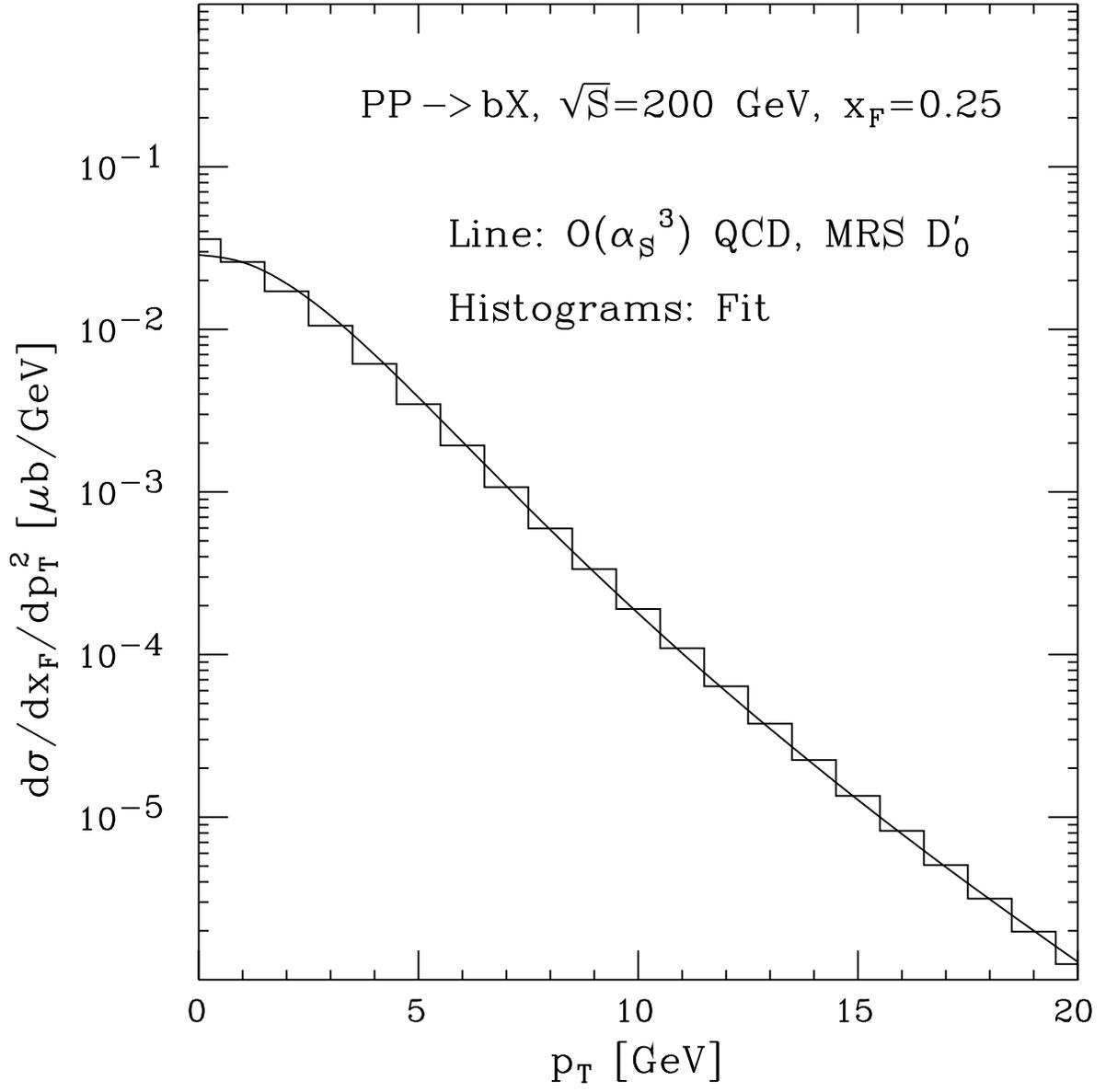
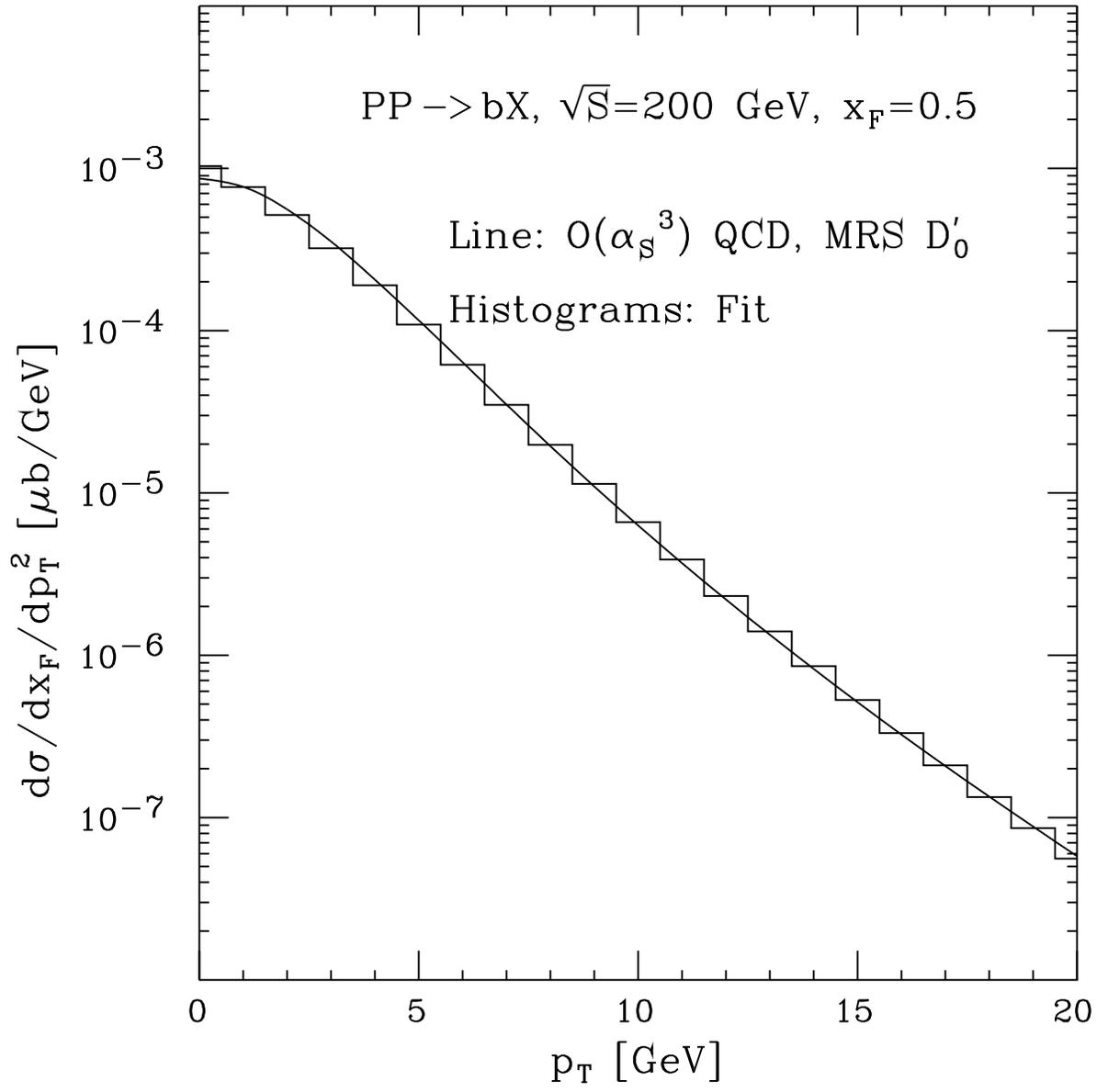
\item[Fig.  3.]
As in Fig. 2 but for the proposed SSC fixed target energy $\sqrt{s}=200$ GeV.

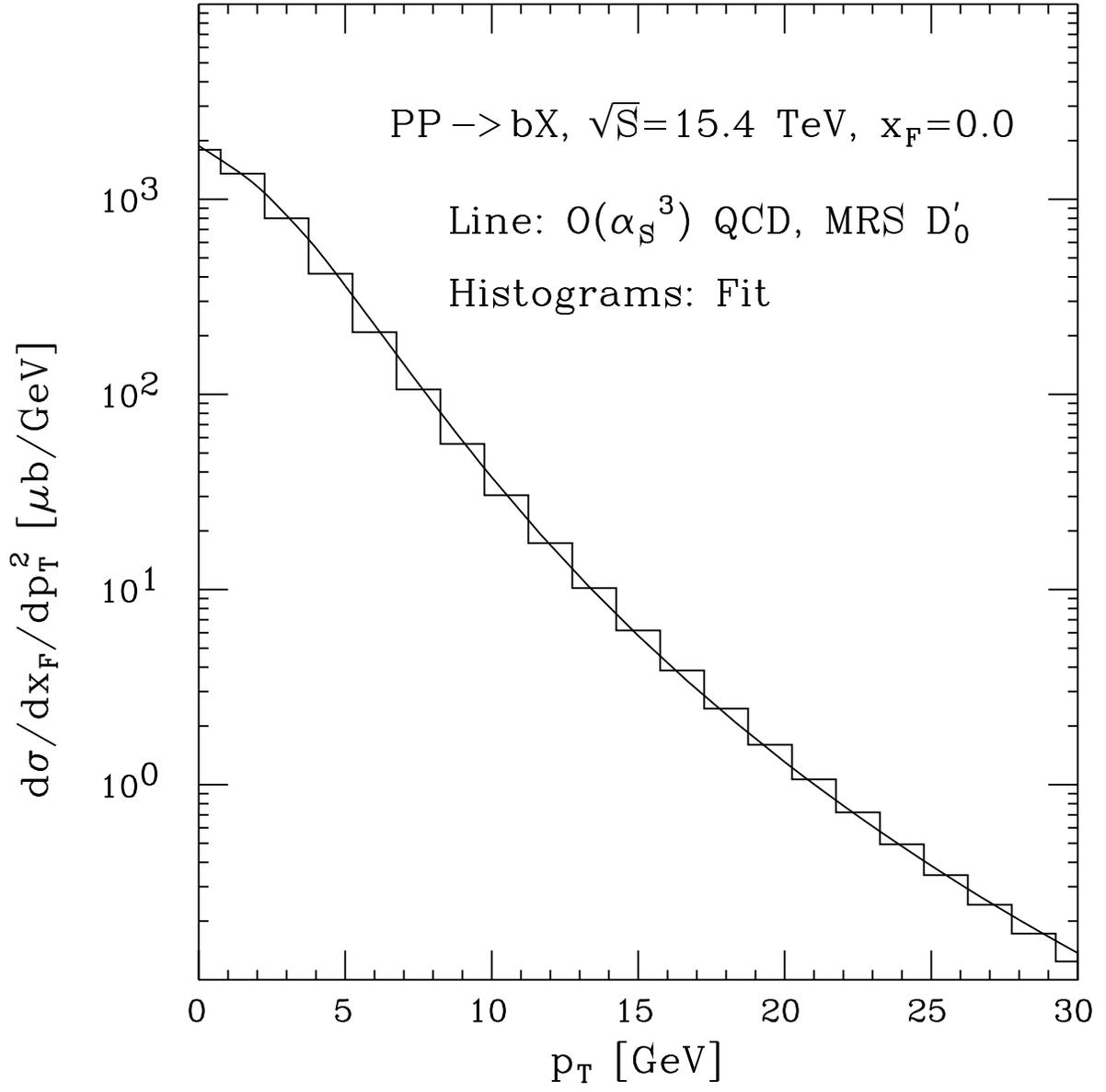
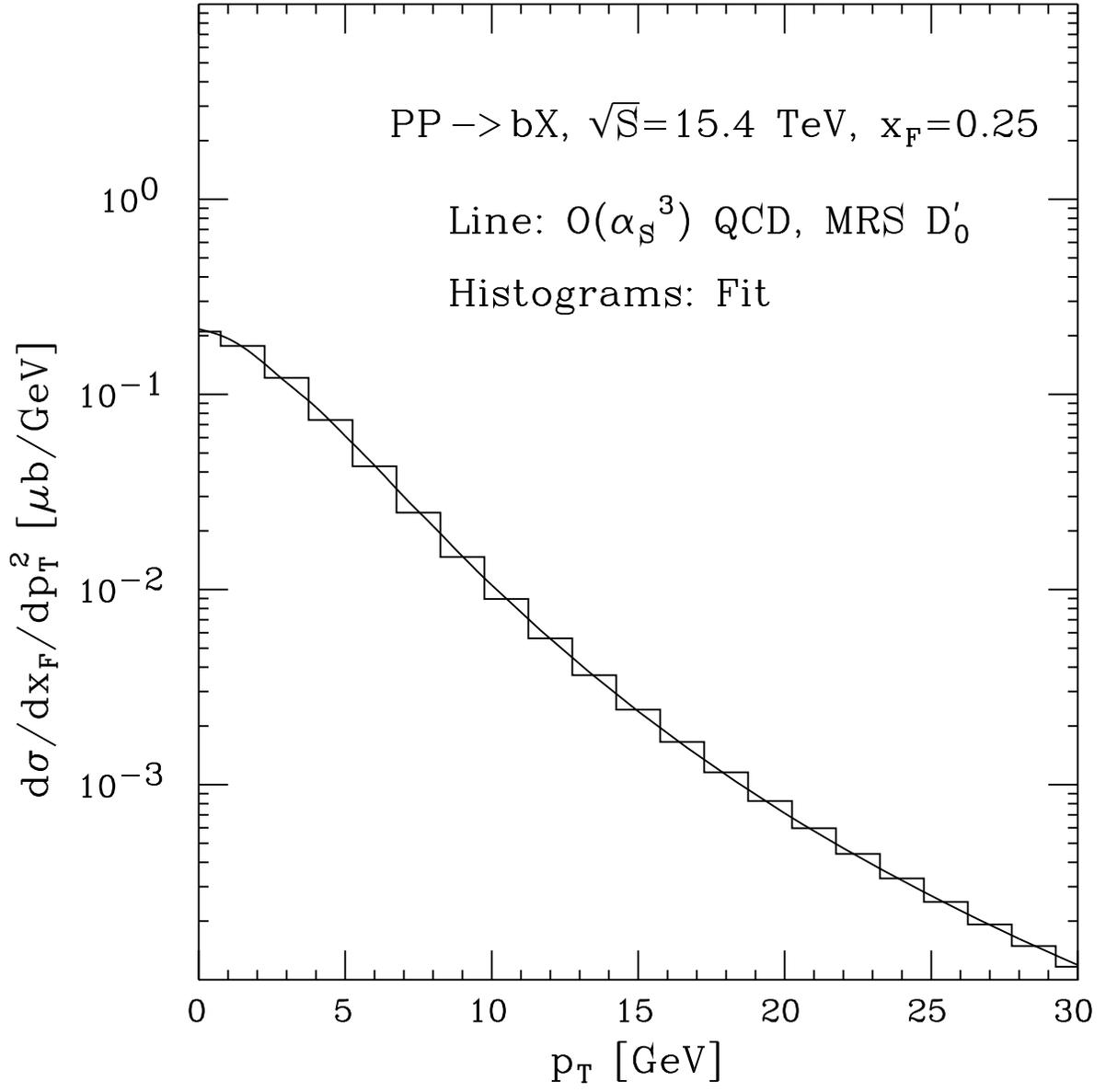
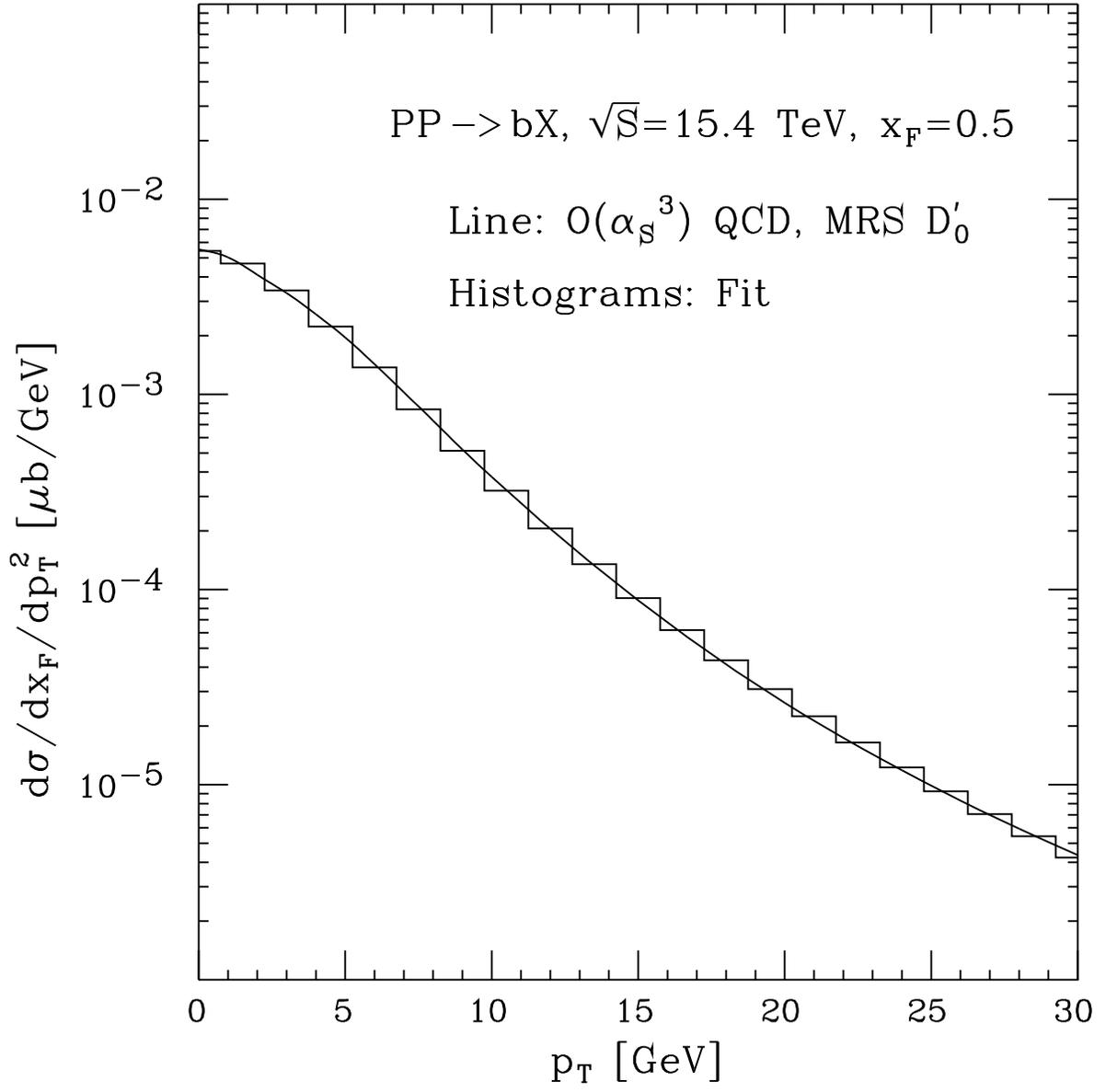
\item[Fig.  4.]
As in Fig. 2 but for the LHC collider energy $\sqrt{s}=15.4~$TeV.

\end{description}
\end{document}